\documentclass[12pt]{article}

\begin{document}

\begin{center}

{\Large {Quark mixing angles and weak CP-violating phase vs quark masses: potential approach}}\footnote{This paper develops results and corrects some mistakes of Preprint arXiv:2210.09780}

\vspace{1,5cm}

{Boris L. Altshuler}\footnote{E-mail addresses: baltshuler@yandex.ru \,\,\,  \&  \,\,\,  altshul@lpi.ru}

\vspace{1cm}

{\it {Theoretical Physics Department, Lebedev Physical
Institute of the Russian Academy of Sciences, \\  53 Leninsky Prospect, Moscow, 119991, Russia}}

\vspace{1,5cm}

\end{center}

{\bf {Abstract:}} 

It is shown that following experimentally viable expressions for quark mixing angles $\theta_{12}$, $\theta_{23}$, $\theta_{13}$ and CP-violating phase $\delta$: $\sin\theta_{12} = \sqrt{m_{d} / |m_{s}|}$, $\sin\theta_{23} = 2 \, |m_{s}| / m_{b}$, $\sin\theta_{13} \approx 2 \, m_{d} / m_{b}$, $\tan\delta = m_{b}^{2} \, m_{c} / 6 \, m_{t} \, m_{s}^{2}$ 
may be derived as stable points of certain 4-th power in $V_{CKM}$ flavor-invariant potentials built with traces of 3x3 quark up and down mass matrices and the Jarlskog invariant. There is no fine-tuning, potentials' dimensionless constants are the integer numbers not above 10.

\vspace{0,5cm}


\vspace{0,5cm}


\newpage

\section{Preliminaries and motivation}

\quad The arbitrariness of fermion masses, mixing angles and CP violating phases of the quark and lepton sectors of Standard Model (SM) plague the SM for decades in spite of the plural efforts to find theory explaining these fundamental numbers, see e.g. \cite{Isidori} - \cite{Feruglio} and references therein. Here the quark sector of SM will be considered which fermion Lagrangian $L_{f}$ \cite{Peskin} we supplement with potential $W(\hat{M}_{u}, \hat{M}_{d})$ depending on 3x3 up and down current quark mass matrices:

\begin{equation}
\label{1}
L_{f} = \bar{\psi}_{u}\hat{\partial}\psi_{u} + \bar{\psi}_{d}\hat{\partial}\psi_{d} + {\bar{\psi}}_{uL}\hat{M}_{u}\psi_{uR} + {\bar{\psi}}_{dL}\hat{M}_{d}\psi_{dR} - W(\hat{M}_{u}, \hat{M}_{d}).
\end{equation}
Potential $W$ will be discussed in Sec. 2. $\hat{M}_{u, d} = (v/\sqrt2) \hat{Y}_{u,d}$ ($v$ is vacuum average of the Higgs field, $\hat{Y}_{u,d}$ are dimensionless matrices which elements are so called Yukawa couplings). In Eq. (\ref{1}) SM gauge fields are omitted, and also are omitted generation (flavor) indices of spinor fields ($\psi_{u} = u, c, t$ quarks; $\psi_{d} = d, s, b$ quarks) and of their mass matrices. 

It is well known \cite{Peskin} that with proper unitary transformations of the left and right Fermi fields $\hat{M}_{u,d}$ may be made hermitian and diagonalized: $\hat{M}_{u} = U_{u}\hat{m}_{u}U^{\dag}_{u}$, $\hat{M}_{d} = U_{d}\hat{m}_{d}U^{\dag}_{d}$; $\hat{m}_{u} = {\rm{diag}}(m_{ui}) = {\rm{diag}}(m_{u}, m_{c}, m_{t})$; $\hat{m}_{d} = {\rm{diag}}(m_{dj}) = {\rm{diag}}(m_{d}, m_{s}, m_{b})$; $i,j = 1,2,3$; $m_{f}$ ($f = u, c, t, d, s, b$) are corresponding six current quark masses; these mass eigenvalues may be positive or negative. In case matrices $\hat{M}_{u,d}$ can't be diagonalized with one and the same unitary transformation the vertex of quarks' interaction with charged gauge bosons $W^{\pm}$ includes, in the SM mass eigenvalues basis, the flavor-mixing unitary Cabibbo-Kobayashi-Maskawa matrix $\hat{V}_{CKM} = U^{\dag}_{u}U_{d}$ \cite{Peskin} depending on three mixing angles $\theta_{12}$, $\theta_{23}$, $\theta_{13}$ and the weak CP violating phase $\delta$.

Thus two Hermitian matrices $\hat{M}_{u,d}$ hide in ten experimentally observed constants of the SM quark sector: masses of six quarks and four named above parameters of CKM matrix. The updated values of the current quarks masses (at scale 2 GeV; errors are shown in brackets) are presented in \cite{masses}: $m_{u}$ = 2.16(38) MeV, $m_{c}$ = 1.27(2) GeV, $m_{t}$ = 172.4(7) GeV; $m_{d}$ = 4.67(32) MeV, $m_{s}$ = 0.093(8) GeV, $m_{b}$ = 4.18(2) GeV (if the eigenvalues of matrices $\hat{M}_{u,d}$ are negative, then these numbers refer to the modules of the eigenvalues). Three CKM-angles and CP-violating phase are also observed with high accuracy \cite{angles}. In the standard PDG parametrization $\hat{V}_{CKM}$ ($\equiv \hat{V}$ below) looks as:

\begin{equation}
\label{2}
\begin{array}{c}
\hat{V} = \hat{V}_{23} \times \hat{V}_{13} \times \hat{V}_{12} = \\
\\
= \left(\begin{array}{ccc}
1 & 0 & 0 \\
0 & c_{23} & s_{23} \\
0 & -s_{23} & c_{23}
\end{array}\right) \times
\left(\begin{array}{ccc}
c_{13} & 0 & s_{13} e^{-i\delta} \\
0 & 1 & 0 \\
- s_{13} e^{i\delta} & 0 & c_{13}
\end{array}\right) \times
\left(\begin{array}{ccc}
c_{12} & s_{12} & 0 \\
-s_{12} & c_{12} & 0 \\
0 & 0 & 1
\end{array}\right) = \\
\\
= \left(\begin{array}{ccc}
c_{12}c_{13} & s_{12}c_{13} & s_{13} e^{-i\delta} \\
- s_{12} c_{23} - c_{12} s_{23}s_{13}e^{i\delta} & c_{12} c_{23} - s_{12} s_{23}s_{13}e^{i\delta} & s_{23}c_{13} \\ 
s_{12} s_{23} - c_{12} c_{23}s_{13}e^{i\delta}  & - c_{12} s_{23} - s_{12} c_{23}s_{13}e^{i\delta}  & c_{23} c_{13}  \nonumber
\end{array}\right),
\end{array}
\end{equation}
here $s_{12} = \sin\theta_{12}$, $c_{12} = \cos\theta_{12}$ etc.

The observed values of these ten constants obey hierarchical structure. In particular, there is the so far unexplained correlation (which is the main focus of the present paper) of hierarchy of quark masses and hierarchy of the CKM matrix mixing angles. These hierarchies and their correlation are immediately visible if the CKM angles and ratios of quark masses are expressed with one and the same Wolfenstein parameter $\lambda$ \cite{Wolfenstein} (the observed value of the weak CP violating phase $\delta$ is also given below):

\begin{equation}
\label{3}
\begin{array}{c}
\lambda = s_{12} = 0.2253(7); \, \, \, \, \, s_{23} = 4.080(14)\cdot 10^{-2} = 0.81 \, \lambda^{2}; \\
s_{13} = 3.82(20) \cdot 10^{-3} = 0.34 \, \lambda^{3}; \, \delta = 69(4)^{\circ}; \, \tan\delta = 2.6(6); \\
\frac{m_{u}}{m_{c}} = 1.70(40) \cdot 10^{-3} = 0.65 \, \lambda^{4}; \qquad \frac{m_{d}}{m_{s}} = 0.050(7) = \lambda^{2}; \qquad \qquad \quad \\
\frac{m_{c}}{m_{t}} = 0.747(12) \cdot 10^{-2} = 2.9 \, \lambda^{4}; \qquad \frac{m_{s}}{m_{b}} = 2.22(25) \cdot 10^{-2} = 0.44 \, \lambda^{2}; \, \, \\
\frac{m_{u}}{m_{t}} = 1.26(20) \cdot 10^{-5} = 1.88 \, \lambda^{8}; \qquad \frac{m_{d}}{m_{b}} = 1.12(8) \cdot 10^{-3} = 0.44 \, \lambda^{4}; \, \, \, \, \,
\end{array}
\end{equation}
the values of quark masses given above are used here; values of $V_{CKM}$ parameters see in \cite{angles}.

The task to explain the origin of these numbers and their specific hierarchy and tuning is called "flavor puzzle". The part of this "puzzle" is to find the well grounded expressions for the CKM parameters through the ratios of quark masses. This paper follows the approach of \cite{Anselm}, \cite{Alonso1} where six quark masses are treated as fixed constants known from experiment, whereas values of the CKM parameters should be determined by the minimization of certain effective potential depending on $V_{CKM}$. The origin of this potential, as it is stated in \cite{Anselm}, may be quantum vacuum diagrams with the charged gauge bosons or charged Higgs exchange which have $\hat{V}_{CKM}$ in their vertexes. However authors of \cite{Anselm}, \cite{Alonso1} build the looked for potential with the simplest $V_{CKM}^{2}$ combination ${\rm{Tr}}{\hat{M}}^{\dag}_{u}M_{u}{\hat{M}}^{\dag}_{d}M_{d}$ which by itself gives trivial extremal values of the CKM angles. To get the sensible results the additional terms of different powers of the named combination and fine-tuning of constants are used in \cite{Anselm}, \cite{Alonso1}.

In the present paper potential $W$ (\ref{1}) is built with flavor invariants including 0-, 2-, 4-powers of $\hat{V}_{CKM}$, the 4-$\hat{V}_{CKM}$ terms may originate from the 4-vertexes quantum vacuum diagrams (like it was noted in \cite{Anselm} for 2-$\hat{V}_{CKM}$ terms of potential). This permits to obtain for the CKM angles the  Mexican hat potential with stable extrema. Also we postulate potential $W$ being homogeneous in mass matrices and symmetric under exchange of $\hat{M}_{u}$ and $\hat{M}_{d}$.  Thus there are no fine-tuned dimensional constants in the potential introduced below, more of that: dimensionless constants before the different terms of potential are the integer numbers not above 10. And this allows to obtain the mass-mixing relations compatible with experimental data. It may be surprising.

A significant feature of our approach is that we initially consider the Hermitian mass matrices of quarks $\hat{M}_{u}$, $\hat{M}_{d}$ (which, as noted above, is always possible \cite{Peskin}) and build the potential $W$ from them. While in \cite{Anselm}, \cite{Alonso1} and in many other works potential is constructed from the Hermitian quadratic combinations ${\hat{M}}^{\dag}_{u}M_{u}$ and ${\hat{M}}^{\dag}_{d}M_{d}$ where mass matrices are arbitrary. This difference is essential in obtaining of the numerical predictions below.

The idea of spontaneous flavor symmetry breaking (SFSB), when Yukawa couplings are considered as dynamical fields ("spurions", "flavons") and observed values of fermion masses and of mixing angles are these fields' VEVs to be determined as the extremal points of some flavor invariant potential looks quite attractive. Actually, the idea is rather old \cite{Cabibbo}, \cite{Gatto}, it was developed later by a number of authors where possibilities to explain the origin of hierarchies of fermion masses and mixing angles in both - SM quark and lepton sectors were studied, see \cite{Isidori}, \cite{Feruglio}, \cite{Anselm}, \cite{Alonso1}, \cite{Ambrosio} - \cite{Nardi}, and references therein. 

The general difficulty of the SFSB approach are the Goldstone bosons that would result from the spontaneous breaking of a continuous global flavor symmetry. Gauging of flavor symmetry resolves this problem but in turn tends to induce dangerous flavor–changing neutral currents mediated by the new gauge bosons. This problem may be also circumvented, see in \cite{Albrecht}, \cite{difficulties} and later references in \cite{Alonso}. We shall not touch these issues here, but considering $V_{CKM}$ parameters as numbers just try to build potential introduced in (\ref{1}) which extremization must give experimentally viable dependencies of these parameters on the observed quarks mass ratios (\ref{3}).

Some of results below are obtained in the two generations approximation. This approximation for light quarks (when only $s_{12} \ne 0$ in $V_{CKM}$ (\ref{2})) was discussed in detail in \cite{Alonso1} where, in order to obtain the correct value of the Cabibbo angle as an extremum of a certain potential, the fine-tuning of three parameters of order from $10^{-2}$ to $10^{-10}$ is applied. In the present paper the known relation (\ref{6}) between sinus of Cabibbo angle and ratio of the light quark masses is received in the two generations approximation without any fine-tuning.

This paper considers the quark sector of the Standard Model. Lepton sector is more involved because of the Majorana nature of neutrinos, because there are not two but three Yukawa matrices, one additional for the heavy right-handed neutrinos; also there are no visible correlation of lepton masses and large mixing angles of the 
Pontecorvo - Maki - Nakagawa - Sakata matrix, see \cite{Alonso}, \cite{Alonso2} and references therein. 

\section{Potential approach: calculation of $V_{CKM}$ angles and CP-violating phase}

\qquad Jarlskog determinant of the commutator of $\hat{M}_{u}$ and $\hat{M}_{d}$ matrices (or of their squares) is the most famous flavor invariant \cite{Jarlskog}; its non-zero value signals the CP violation. This determinant may be expressed through certain trace of mass matrices \cite{Branco}:

\begin{equation}
\label{4}
\begin{array}{c} 
{\rm{Det}}\{[\hat{M}_{u},\hat{M}_{d}]\} = \frac{1}{3}\,{\rm{Tr}}\{[\hat{M}_{u},\hat{M}_{d}]^{3}\} = 2i \, {\rm{Im}} {\rm{Tr}}\{\hat{M}_{u}\hat{M}_{d}\hat{M}_{u}^{2}\hat{M}_{d}^{2}\} = \\
(m_{c} - m_{u})\,(m_{t} - m_{u})\,(m_{t} - m_{c})\, (m_{s} - m_{d})\,(m_{b} - m_{d})\,(m_{b} - m_{s})\,2 i J; \\
J = s_{12}c_{12}s_{23}c_{23}s_{13}c_{13}^{2} \sin\delta = 3.18(0.15) \cdot 10^{-5}.
\end{array}
\end{equation}
Last figure is an experimentally observed value of $J$. In what follows we shall use real part of ${\rm{Tr}}\{\hat{M}_{u}\hat{M}_{d}\hat{M}_{u}^{2}\hat{M}_{d}^{2}\}$ (\ref{4}) as well as some other traces that include zero, second and fourth powers of the CKM matrix. 

In \cite{invariants} all independent traces of two 3x3 matrices are listed which number is limited according to the Caylay-Hamilton theorem. In \cite{Bingrong1}, \cite{Lu} the flavor invariants in leptonic sector, which is essentially more complicated than the quark sector, are investigated. Papers \cite{Bingrong2} consider flavor invariants as the sources of CP violation in the canonical seesaw model for neutrino masses and that in the seesaw effective field theory; it is shown that all the physical parameters can be extracted using the primary invariants (see also second work \cite{Jarlskog} and references therein). 

Developing the flavor-invariant approach of papers \cite{Jarlskog} - \cite{Bingrong2} the explicit dependences of some flavor invariants on quark masses and CKM matrix parameters are derived in the present paper. The building blocks for potential $W(\hat{M}_{u}, \hat{M}_{d})$ (\ref{1}) are the flavor invariant traces ${\rm{Tr}}\{\hat{M}_{u}^{n_{1}}\} = \sum_{i=1}^{3}m_{ui}^{n_{1}}$, ${\rm{Tr}}\{\hat{M}_{d}^{n_{2}}\} = \sum_{j=1}^{3}m_{dj}^{n_{2}}$, and "mixing" traces  of type: 

$$
{\rm{Tr}}\{\hat{M}_{u}^{n_{1}}\hat{M}_{d}^{n_{2}}\} = {\rm{Tr}}\{\hat{m}_{u}^{n_{1}}\hat{V}\hat{m}_{d}^{n_{2}}\hat{V}^{\dag}\},
$$
$$
{\rm{Tr}}\{\hat{M}_{u}^{n_{1}}\hat{M}_{d}^{n_{2}}\hat{M}_{u}^{n_{3}}\hat{M}_{d}^{n_{4}}\} = {\rm{Tr}}\{\hat{m}_{u}^{n_{1}}\hat{V}\hat{m}_{d}^{n_{2}}\hat{V}^{\dag}\hat{m}_{u}^{n_{3}}\hat{V}\hat{m}_{d}^{n_{4}}\hat{V}^{\dag}\}.
$$

General expressions for the last two "mixing" traces as functions of moduli of the "angle" elements of unitary matrix $\hat{V}$ are given in Appendix for arbitrary diagonal matrices that alternate with $\hat{V}$ and $\hat{V}^{\dag}$.

As it was noted in the Introduction, we shall consider potential $W$ in (\ref{1}) to be symmetric with exchange $\hat{M}_{u} \leftrightarrow \hat{M}_{d}$ and homogeneous in masses; thus there will be no ad hoc dimensional constants which influence the final quantitative results of the paper. The looked for mixing angles will depend on mass ratios (\ref{3}); more of that: because of strong inequality of the "up" and "down" mass ratios the main contributions in expressions for mixing angles come only from the "down" ratios $m_{d} / m_{s}$, $m_{s} / m_{b}$, $m_{d} / m_{b}$.

\vspace{0,5cm}

\quad {\large {\bf{2.1. Calculation of $s_{12}$.}}}

\vspace{0,5cm}

In calculation of the mixing of two light generations we follow the two generations approach when relatively small, as compared to $s_{12}$, mixing angles $\theta_{23}$, $\theta_{13}$ of third and lighter generations are put to zero. Let us look at the 4-th power and homogeneous in masses potential when $\theta_{23}$, $\theta_{13}$ in (\ref{2}) are zero and only $\theta_{12} \ne 0$ (definition of $\hat{V}_{12}$ see in (\ref{2})):

\begin{equation}
\label{5}
\begin{array}{c}
W^{(4)}(s_{12}) = {\rm{Re}} \, \large[{\rm{Tr}}\{\hat{M}_{u}\hat{M}_{d}\hat{M}_{u}\hat{M}_{d}\} - 2 \, {\rm{Tr}}\{\hat{M}_{u}^{2}\hat{M}_{d}^{2}\}\large] = \nonumber \\ 
= {\rm{Re}} \, \large[{\rm{Tr}}\{\hat{m}_{u}\hat{V}_{12}\hat{m}_{d}\hat{V}_{12}^{\dag}\hat{m}_{u}\hat{V}_{12}\hat{m}_{d}\hat{V}_{12}^{\dag}\}
- 2 \, {\rm{Tr}}\{\hat{m}_{u}^{2}\hat{V}_{12}\hat{m}_{d}^{2}\hat{V}_{12}^{\dag}\}\large] \approx \\
\approx - \sum_{i=1}^{3}m_{ui}^{2}m_{di}^{2} + m_{c}^{2} [m_{s}^{2}s_{12}^{4} + 2 \, m_{s}m_{d}s_{12}^{2}]. \\
\end{array}
\end{equation}

The last approximate expression in the chain of equations (\ref{5}) is received from the general formulas (A4), (A6) of Appendix, where only contributions of the largest mass ratios were taken into account. It is seen that for the negative sign of the second eigenvalue $m_{s}$ of mass matrix $\hat{M}_{d}$ potential (\ref{5}) as a function of $s_{12} $ has a Mexican hat form with stable nonzero minimum at:

\begin{equation}
\label{6}
s_{12} = \bar{s}_{12} = \sqrt{\frac{m_{d}}{|m_{s}|}} = s_{12}^{exp}.
\end{equation}
Whereas an unstable extremum of this potential, at $s_{12} = 0$, corresponds to $\hat{V}_{CKM} = \hat{1}$. Formula (\ref{6}) for Cabibbo angle is known for decades \cite{Cabibbo}, \cite{Gatto}, but, as to our knowledge, it was not derived earlier from the extremization of potential of type (\ref{5}).

The analogous calculation of potential (\ref{5}) in case $s_{23} \ne 0$, $s_{12} = s_{13} = 0$ gives for $s_{23}$ the similar to (\ref{5}) 4-th power potential which flavor braking minimum at $\bar{s}_{23} = \sqrt{|m_{s}|/m_{b}}$ is in gross contradiction with experimental data (\ref{3}). Also nothing good will come out from attempt to calculate $s_{13}$ from potential (\ref{5}).

\vspace{0,5cm}
 \quad {\large {\bf{2.2. Calculation of $s_{23}$, $s_{13}$.}}}

\vspace{0,5cm}

To get the sensible extremal values of $s_{23}$ and $s_{13}$ more complicated 8-th power in masses potential $W$ (\ref{1}) must be considered:

\begin{equation}
\label{7}
\begin{array}{c}
W = \kappa \, W^{(8)} = \kappa \, {\rm{Re}} \, \large[c_{1}W_{1} + c_{2} W_{2} + c_{3} W_{3} + c_{4} W_{4}\large], \\
W_{1} = ({\rm{Tr}}\hat{M}_{u})^{2} \cdot ({\rm{Tr}}\hat{M}_{d})^{2} \cdot [{\rm{Tr}}\{\hat{M}_{u}\hat{M}_{d}\hat{M}_{u}\hat{M}_{d}\} - 2 \, {\rm{Tr}}\{\hat{M}_{u}^{2}\hat{M}_{d}^{2}\}]; \nonumber \\
W_{2} = {\rm{Tr}}\hat{M}_{u} \cdot {\rm{Tr}}\hat{M}_{d} \cdot [{\rm{Tr}}\{\hat{M}_{u}\hat{M}_{d}\hat{M}_{u}^{2}\hat{M}_{d}^{2}\} - 2 \, {\rm{Tr}}\{\hat{M}_{u}^{3}\hat{M}_{d}^{3}\}]; \nonumber \\
W_{3} = {\rm{Tr}}\{\hat{M}_{u}\hat{M}_{d}\hat{M}_{u}^{3}\hat{M}_{d}^{3}\} - 2 \, {\rm{Tr}}\{\hat{M}_{u}^{4}\hat{M}_{d}^{4}\}; \nonumber \\
W_{4} = {\rm{Tr}}\{\hat{M}_{u}^{2}\hat{M}_{d}^{2}\hat{M}_{u}^{2}\hat{M}_{d}^{2}\} - 2 \, {\rm{Tr}}\{\hat{M}_{u}^{4}\hat{M}_{d}^{4}\}, \nonumber
\end{array}
\end{equation}
$\kappa$ is a dimensional constant irrelevant here, $c_{1,2,3,4}$ are dimensionless constants determined below.

With use of general formulas of Appendix and leaving only the largest terms in the hierarchy of mass ratios, we present below the expressions for potential (\ref{7}) in three "two generations" approximations ($W^{(8)}_{0}$ is an unessential here value of potential (\ref{7}) at $\hat{V} = \hat{1}$).

1) $s_{12} \ne 0$, $s_{23} = s_{13} = 0$ in (\ref{2}).

In this case only "angle" parameter $p$ (A1) is not equal to zero in (A3), (A5) and the main term of (\ref{7}) is $W_{1}$ which is potential (\ref{5}) multiplied by $m_{t}^{2} \, m_{b}^{2}$:

\begin{equation}
\label{8}
W^{(8)}(s_{12}) \approx W^{(8)}_{0} + m_{t}^{2} \, m_{b}^{2} \, m_{c}^{2} \cdot c_{1} \, [m_{s}^{2} \, s_{12}^{4} - 2\,|m_{s}| \, m_{d} \, s_{12}^{2}].
\end{equation}
$c_{1} > 0$ is a necessary condition for potential (\ref{8}) to have minimum at $s_{12} = {\bar{s}}_{12}$ (\ref{6}).

2) $s_{23} \ne 0$, $s_{12} = s_{13} = 0$.

We have from (\ref{7}), (A3), (A5) with account of (A2), (A4), (A7):

\begin{equation}
\label{9}
\begin{array}{c}
W^{(8)}(s_{23}) \approx W^{(8)}_{0} + m_{t}^{4} \, m_{b}^{2} \cdot \{s_{23}^{4} \, m_{b}^{2} \, \sum_{k=1}^{4}c_{k} + \\
+ s_{23}^{2}\, [|m_{s}| \, m_{b} \, (2c_{1} + c_{2} + c_{3}) - \, m_{s}^{2}\, (2c_{1} - c_{2} -2 c_{4})]\}.
\end{array}
\end{equation}
This expression is easily derived taking into account that in case $s_{12} = s_{13} = 0$ only "angular" parameter $s$ (see (A1), (A2)) is non-zero in (A3), (A5). 

Condition

\begin{equation}
\label{10}
2c_{1} + c_{2} + c_{3} = 0
\end{equation}
is necessary to exclude in (\ref{9}) a linear in $m_{s}$ "dangerous" term giving a physically unacceptable extremal value of $s_{23}$. Thus, with account of (\ref{10}), minimum of potential (\ref{9}) is reached at:

\begin{equation}
\label{11}
s_{23} = \bar{s}_{23} = \sqrt{\frac{2c_{1} - c_{2} -2 c_{4}}{2\, \sum_{k=1}^{4}c_{k}}} \cdot \frac{|m_{s}|}{m_{b}}.
\end{equation}

The fulfillment of the experimentally viable (see in (\ref{3})) relation

\begin{equation}
\label{12}
\bar{s}_{23} = 2 \, \frac{|m_{s}|}{m_{b}} = s_{23}^{exp}
\end{equation}

demands in (\ref{11}):

\begin{equation}
\label{13}
\frac{2c_{1} - c_{2} -2 c_{4}}{2\, \sum_{k=1}^{4}c_{k}} = 4.
\end{equation}

3) $s_{13} \ne 0$, $s_{12} = s_{23} = 0$.

In this case parameters (A1) are equal: $p = q = r = s = s_{13}^{2}$ and according to (A4), (A8) main terms of potential (\ref{7}) look as:

\begin{equation}
\label{14}
\begin{array}{c}
W^{(8)}(s_{13}) \approx W^{(8)}_{0} + m_{t}^{4} \, m_{b}^{2} \cdot \{s_{13}^{4} \, m_{b}^{2} \, \sum_{k=1}^{4}c_{k} + \\
+ s_{13}^{2}\, [m_{d} \, m_{b} \, (2c_{1} + c_{2} + c_{3}) - \, m_{d}^{2}\, (2c_{1} - c_{2} -2 c_{4})]\}.
\end{array}
\end{equation}

Taking into account conditions (\ref{10}), (\ref{13}) we see that value of $s_{13}$ corresponding to the minimum of potential (\ref{14}) is given by formula similar to (\ref{11}), (\ref{12}) as it could be expected:

\begin{equation}
\label{15}
s_{13} = \bar{s}_{13} = \sqrt{\frac{2c_{1} - c_{2} -2 c_{4}}{2\, \sum_{k=1}^{4}c_{k}}} \cdot \frac{m_{d}}{m_{b}} = 2 \, \frac{m_{d}}{m_{b}} \approx s_{13}^{exp}.
\end{equation}
Substitution here of the experimental value of ratio $m_{d}/m_{b}$ (\ref{3}) gives $\bar{s}_{13} = 2.24(16) \cdot 10^{-3}$ which is 1.7 times less than the observed value of $s_{13} = 3.82(20) \cdot 10^{-3}$ (see (\ref{3})).

The set of integer constants $c_{k}$ in (\ref{7}) that satisfy conditions (\ref{10}), (\ref{13}) is for example: $c_{1} = 3$, $c_{2} = -10$, $c_{3} = 4$, $c_{4} = 4$. Then potential (\ref{7}) looks as:

\begin{equation}
\label{16}
W^{(8)} = {\rm{Re}} \, \large[3 \, W_{1} - 10 \, W_{2} + 4 \, W_{3} + 4 \, W_{4}\large].
\end{equation}

Important note: calculation of the exact potential (\ref{16}) shows that approximate expression (\ref{9}) for $W^{(8)}(s_{23})$ is valid without simplifying assumption $s_{12} = s_{13} = 0$; the reason is the strong hierarchy of quark masses $m_{t} \gg m_{u}, m_{c}$; $m_{b} \gg m_{d}, m_{s}$. Thus expression (\ref{11}) for the extremal value $\bar{s}_{23}$ does not need the two generations approximation. Unfortunately, the same cannot be said about the results (\ref{6}), (\ref{15}) for the extremal values $\bar{s}_{12}$, $\bar{s}_{13}$ because the "crossing terms" in exact potential destroy these results. Thus potential $W^{(8)}(s_{12}, s_{23}, s_{13})$ (\ref{16}) possesses exact minimum at the value of $s_{23}$ (\ref{12}) and two conditional minima (\ref{6}), (\ref{15}) in variables $s_{12}$, $s_{13}$.

\vspace{0,5cm}
 \quad {\large {\bf{2.3. Calculation of CP violating phase $\delta$}}}.

\vspace{0,5cm}

Let us pay attention that Jarlskog determinant (\ref{4}) contains $\sin\delta$ whereas parameter $r = |V_{31}|^{2}$ (A2) contains $\cos\delta$ (other parameters $p, q, s$ (A1), (A2) do not depend on $\delta$) - both with one and the same extremely small multiplier, and that the extremization of expression $a\, \cos\delta + b \, \sin\delta$ gives $\tan\delta = b/a$. Thus certain sum of Jarlskog determinant (\ref{4}) and linear in $r$ terms of potential $W(\hat{M}_{u}, \hat{M}_{d})$ may hopefully give the wishful extremal value of $\delta$. Here is the simplest homogeneous in masses potential which minimum is placed at the phenomenologically favorable value of CP violating phase $\delta$:

\begin{equation}
\label{17}
\begin{array}{c}
W(\delta) = i \,{\rm{Det}}\{[\hat{M}_{u},\hat{M}_{d}]\} - 6 \cdot {\rm{Tr}}\{\hat{M}_{u}^{3}\hat{M}_{d}^{3}\} \approx \\
\\
\approx 2\, s_{12}c_{12}s_{23}c_{23}s_{13}\, m_{t}^{2}m_{s} \, [- c_{13}^{2}\, m_{b}^{2} \, m_{c} \, \sin\delta - 6 \, m_{t} \, m_{s}^{2} \, \cos\delta]
\end{array}
\end{equation}
(only the main terms depending on $\delta$ are left here in the RHS which is calculated with account of (\ref{4}), (A3) and expression for $r$ in (A2)).

Potential (\ref{17}) is minimal at (we put $c_{13}^{2} = 1$)

\begin{equation}
\label{18}
\tan\delta = \tan\bar{\delta} = \frac{m_{b}^{2}m_{c}}{6 \, m_{t}m_{s}^{2}} = 2.5(6),.
\end{equation}
which is in a good agreement with experimental value of $\tan\delta$ shown in (\ref{3}).

\section{Conclusion}

\quad The results of this work can probably be considered purely methodical: the work "proves the existence theorem", that is, it demonstrates the possibility of choosing such potentials (\ref{1}) whose stable points (\ref{6}), (\ref{12}), (\ref{15}), (\ref{18}) correspond to the observed values of the mixing angles and the CP violating phase of the Cabibbo-Kobayashi-Maskawa matrix. We hope that it may stimulate further research in this area, including possible rationale for the choice of potential and looking beyond the two-generations approximation of potential for mixing angles $s_{12}$, $s_{13}$.

To study the applicabiity of results of this paper to the lepton sector of Standard Model in context of potential approach in \cite{Alonso}, \cite{Alonso2} and in similar papers is an another challenge for future. Some difficulties on this way are named at the end of the Introduction.

\section*{Acknowledgements} Author is grateful to Mikhail Vysotsky for permanent consultations and to Valery Rubakov, who died untimely in October 2022, for the inspiring discussions on the ABC of Standard Model and its problems.

\section*{Appendix}

\qquad Formulas below present the expressions of the alternating "chain" products of arbitrary diagonal 3x3 matrices $\hat{\alpha} = {\rm{diag}}(\alpha_{1}, \alpha_{2}, \alpha_{3})$ etc. with arbitrary 3x3 unitary matrices $\hat{V}$ and $\hat{V}^{\dag}$ through the differences of eigenvalues $\alpha_{12} = \alpha_{1} - \alpha_{2}$ etc., and modules of the "angle" elements $v_{11}, v_{33}, v_{13}, v_{31}$ of matrix $\hat{V}$:

$$
p = 1 - |v_{11}|^{2}; \quad s = 1 - |v_{33}|^{2}; \quad q = |v_{13}|^{2}; \quad r = |v_{31}|^{2}. \eqno (A1)
$$

For $\hat{V}_{CKM}$ (\ref{2}):

$$
\begin{array}{c}
p = s_{12}^{2} + s_{13}^{2} - s_{12}^{2} \, s_{13}^{2}; \quad s = s_{23}^{2} + s_{13}^{2} - s_{23}^{2} \, s_{13}^{2}; \quad q = s_{13}^{2};   \\
r = s_{13}^{2} + s_{12}^{2} \, s_{23}^{2} - s_{12}^{2} \, s_{13}^{2} - s_{23}^{2} \, s_{13}^{2} + (s_{12}\, s_{23} \, s_{13})^{2} - \\ 
- 2 \, s_{12}\, c_{12} \, s_{23} \, c_{23} \, s_{13} \, \cos\delta.
\end{array}  \eqno  (A2)
$$

Thus for the "chain" trace with two matrices $\hat{V}$, $\hat{V}^{\dag}$ it is obtained:

$$
{\rm{Tr}}\{\hat{\alpha}\hat{V}\hat{\beta}\hat{V}^{\dag}\} = \sum_{i=1}^{3}\alpha_{i}\beta_{i} - [p \, \alpha_{12}\beta_{12} + q \, \alpha_{12}\beta_{23} + r \, \alpha_{23}\beta_{12} + s \, \alpha_{23}\beta_{23}]. \eqno  (A3)
$$

For three specific cases considered above we have with account of (A2) ($\hat{V}_{12}$, $\hat{V}_{23}$, $\hat{V}_{13}$ are defined in (\ref{2})):

$$
\begin{array}{c}
1) \qquad s_{12} \ne 0: \,\, \, \,  {\rm{Tr}}\{\hat{\alpha}\hat{V}_{12}\hat{\beta}\hat{V}_{12}^{\dag}\} = \sum_{i=1}^{3}\alpha_{i}\beta_{i} - s_{12}^{2} \, \alpha_{12}\beta_{12}; \\
2) \qquad s_{23} \ne 0: \,\, \, \, {\rm{Tr}}\{\hat{\alpha}\hat{V}_{23}\hat{\beta}\hat{V}_{23}^{\dag}\} = \sum_{i=1}^{3}\alpha_{i}\beta_{i} - s_{23}^{2} \, \alpha_{23}\beta_{23}; \\
3) \qquad s_{13} \ne 0: \,\, \, \, {\rm{Tr}}\{\hat{\alpha}\hat{V}_{13}\hat{\beta}\hat{V}_{13}^{\dag}\} = \sum_{i=1}^{3}\alpha_{i}\beta_{i} - s_{13}^{2} \, \alpha_{13}\beta_{13}.
\end{array}  \eqno  (A4)
$$

The "chain" trace with four matrices $\hat{V}$, $\hat{V}^{\dag}$ looks as:

$$
\begin{array}{c}
{\rm{Tr}}\{\hat{\alpha}\hat{V}\hat{\beta}\hat{V}^{\dag}\hat{\gamma}\hat{V}\hat{\delta}\hat{V}^{\dag}\} = \sum_{i=1}^{3}\alpha_{i}\beta_{i}\gamma_{i}\delta_{i} + \\  
+ [p \, \alpha_{12}\beta_{12} + q \, \alpha_{12}\beta_{23} + r \, \alpha_{23}\beta_{12} + s \, \alpha_{23}\beta_{23}] \, \cdot \\
\cdot \, [p \, \gamma_{12}\delta_{12} + q \, \gamma_{12}\delta_{23} + r \, \gamma_{23}\delta_{12} + s \, \gamma_{23}\delta_{23}] + \\
+ \frac{1}{2} \, (q\,r - p\, s)\, (\alpha_{13}\gamma_{23} - \alpha_{23}\gamma_{13}) \, (\beta_{13}\delta_{23} - \beta_{23}\delta_{13}) - \\ 
- p \, [\alpha_{12}\gamma_{12}\beta_{12}\delta_{12} + (\alpha_{1}\gamma_{1} - \alpha_{2}\gamma_{2}) \, (\beta_{1}\delta_{1} - \beta_{2}\delta_{2})] - \\ 
- s \, [\alpha_{23}\gamma_{23}\beta_{23}\delta_{23} + (\alpha_{2}\gamma_{2} - \alpha_{3}\gamma_{3}) \, (\beta_{2}\delta_{2} - \beta_{3}\delta_{3})] + \\ 
+ q \, [\alpha_{12}\gamma_{12}\beta_{23}\delta_{23} + (\alpha_{2}\gamma_{2} - \alpha_{1}\gamma_{1}) \, (\beta_{2}\delta_{2} - \beta_{3}\delta_{3}) - \\ 
- \frac{1}{2} \, (\alpha_{12}\gamma_{13} + \alpha_{13}\gamma_{12}) \, (\beta_{13}\delta_{23} + \beta_{23}\delta_{13})] + \\ 
+ r \, [\alpha_{23}\gamma_{23}\beta_{12}\delta_{12} + (\alpha_{2}\gamma_{2} - \alpha_{3}\gamma_{3}) \, (\beta_{2}\delta_{2} - \beta_{1}\delta_{1}) - \\ 
- \frac{1}{2} \, (\alpha_{13}\gamma_{23} + \alpha_{23}\gamma_{13}) \, (\beta_{12}\delta_{13} + \beta_{13}\delta_{12})] + \\ 
+ i J \, (\alpha_{13}\gamma_{23} - \alpha_{23}\gamma_{13}) \, (\beta_{13}\delta_{23} - \beta_{23}\delta_{13}),
\end{array}  \eqno(A5)
$$
where $J$ see in (\ref{4}). It is easy to show that for $\hat{\alpha} = \hat{m}_{u}$, $\hat{\beta} = \hat{m}_{d}$, $\hat{\gamma} = \hat{m}_{u}^{2}$, $\hat{\delta} = \hat{m}_{d}^{2}$ imagery part of trace (A5) given in the last line of (A5) is equal to half of Jarlskog determinant (\ref{4}), as it could be expected.

For three specific cases, with account of (A2), expression (A5) comes to:

1) $s_{12} \ne 0$, $s_{23} = s_{13} = 0$:

$$
\begin{array}{c}
{\rm{Tr}}\{\hat{\alpha}\hat{V}_{12}\hat{\beta}\hat{V}_{12}^{\dag}\hat{\gamma}\hat{V}_{12}\hat{\delta}\hat{V}_{12}^{\dag}\} = \sum_{i=1}^{3}\alpha_{i}\beta_{i}\gamma_{i}\delta_{i} + s_{12}^{4} \, \alpha_{12}\gamma_{12}\beta_{12}\delta_{12} - \\  
- s_{12}^{2} \, [\alpha_{12}\gamma_{12}\beta_{12}\delta_{12} + (\alpha_{1}\gamma_{1} - \alpha_{2}\gamma_{2}) \, (\beta_{1}\delta_{1} - \beta_{2}\delta_{2})].
\end{array} \eqno (A6)
$$

2) $s_{23} \ne 0$, $s_{12} = s_{13} = 0$:

$$
\begin{array}{c}
{\rm{Tr}}\{\hat{\alpha}\hat{V}_{23}\hat{\beta}\hat{V}_{23}^{\dag}\hat{\gamma}\hat{V}_{23}\hat{\delta}\hat{V}_{23}^{\dag}\} = \sum_{i=1}^{3}\alpha_{i}\beta_{i}\gamma_{i}\delta_{i} + s_{23}^{4} \, \alpha_{23}\gamma_{23}\beta_{23}\delta_{23} - \\  
- s_{23}^{2} \, [\alpha_{23}\gamma_{23}\beta_{23}\delta_{23} + (\alpha_{2}\gamma_{2} - \alpha_{3}\gamma_{3}) \, (\beta_{2}\delta_{2} - \beta_{3}\delta_{3})].
\end{array} \eqno (A7)
$$

3) $s_{13} \ne 0$, $s_{12} = s_{23} = 0$:

$$
\begin{array}{c}
{\rm{Tr}}\{\hat{\alpha}\hat{V}_{13}\hat{\beta}\hat{V}_{13}^{\dag}\hat{\gamma}\hat{V}_{13}\hat{\delta}\hat{V}_{13}^{\dag}\} = \sum_{i=1}^{3}\alpha_{i}\beta_{i}\gamma_{i}\delta_{i} + s_{13}^{4} \, \alpha_{13}\gamma_{13}\beta_{13}\delta_{13} - \\  
- s_{13}^{2} \, [\alpha_{13}\gamma_{13}\beta_{13}\delta_{13} + (\alpha_{1}\gamma_{1} - \alpha_{3}\gamma_{3}) \, (\beta_{1}\delta_{1} - \beta_{3}\delta_{3})].
\end{array}  \eqno (A8)
$$

\end{document}